\documentstyle[a4,12pt]{article}
\def\Frac#1#2{\frac{\displaystyle{#1}}{\displaystyle{#2}}}

\begin{document}

\begin{center}
{\Large {\bf VIRTUAL VERSUS REAL NUCLEAR 
COMPTON SCATTERING IN THE $\Delta$(1232) 
REGION}}
\end{center}

\vspace{0.5cm}

\begin{center}
{\Large {  A. Gil, J. A. G\'omez Tejedor and E. Oset. }}
\end{center}

\vspace{0.3cm}

{\small {\it
Departamento de F\'{\i}sica Te\'orica and IFIC, Centro Mixto Universidad
de Valencia - CSIC, 46100 Burjassot (Valencia) Spain.}}

\vspace{0.7cm}

\begin{abstract}

In this paper we calculate the cross section for Virtual Compton 
Scattering off nuclei in the delta resonance region. We also calculate the
background for the process from Coherent Bremsstrahlung
in nuclei and explore the regions where the Virtual Compton 
Scattering cross section dominates. The study also shows that it is possible to
extract the cross section for Real Compton Scattering from the
Virtual Compton one in a wide range of scattering angles.

\end{abstract}


PACs numbers: 24.30.Gd , 25.20.-x , 25.30.-c

\newpage

\section{Introduction}
For the last years, Virtual Compton Scattering (VCS) on the nucleon target
has attracted much interest from different points of view. For the case of
very hard photons it could provide a stringent test of perturbative Quantum
Chromo-Dynamics \cite{FAR}, although large experimental difficulties
appear. Below pion threshold it allows one to measure new electromagnetic 
observables which generalize the usual magnetic and electric polarizabilities
\cite{GUI,AUD,BER}. 

For the nuclear target, Virtual Compton Scattering at intermediate energy
should be a highly advanced method to study nuclear structure, although no
data are available so far, and there is only one theoretical study
\cite{PAV}.

On the other hand, for Real Compton Scattering (RCS)  much more
information is available, both for  nucleon and nuclear targets. In the
case of a nucleon target, RCS scattering has been experimentally investigated
in all energy range: in the low energy limit at Mainz \cite{ZIE}, Saskatoon
\cite{HAL} and Illinois \cite{FED}; in the resonance region at Bonn \cite{JUN}
and Tokyo \cite{WAD,ISH}; and in the deep inelastic region at Cornell
\cite{DEU,SHU} and SLAC \cite{AND,CAL}.

RCS on nuclei has been the subject of several recent theoretical
and experimental articles \cite{SCH}-\cite{AU8}.

In this paper we study the coherent Virtual Compton Scattering 
in nuclei, $A (e, e' \gamma) A$. In this process the electron emits a virtual
photon which is scattered in the nucleus producing a real photon in the
final state while the nucleus remain in its ground state.

Apart from the interest by itself, this process could be a useful tool to
investigate Real Compton Scattering in nuclei
 without the need to produce real photons: when the angle between
the final and initial electron is very small, the intermediate virtual photon
is almost real, and then, we can extract the RCS cross section from the VCS
one. 

Why could it be interesting to investigate RCS from VCS?.
From the experimental point of view, if we analyze
both processes  (see Fig. 1), one can see
that the RCS is a two steps process: First, the electron is scattered in a
nucleus in order to produce a real photon from Bremsstrahlung (this process
is of order $Z^2 \alpha^3$), and second, this real photon is scattered in a second
nucleus in order to produce the RCS (this is an order
$(A \alpha)^2$  process);
in total, this process is of order $Z^2 \alpha^5 A^2$. On the
other hand, the VCS is a one step process: the electron exchanges a virtual
photon with the nucleus, and a real photon is emitted; this process is of
order $\alpha (A \alpha)^2$. Therefore, in principle, the ratio between RCS and
VCS should be around $Z^2 \alpha^2$, smaller than 0.1 for most nuclei.
However, this is still a large overestimate since the involvement of two
targets in the real photon case, limits largely the final counts.
Then, in principle, it should be much more efficient to perform an experiment
for VCS than for RCS. However, from the experimental point of view, dealing 
with electrons comport additional difficulties with respect to photons.
 One of the most 
important difficulties is the large background that one would encounter
coming from the Coherent Bremsstrahlung (CB)
in nuclei: in this process the real
photon is emitted from the electron, not from the nucleus.
Thus, we have also studied this process in order to determine the experimental
accessible regions, where the signal for the VCS is clearly visible.

We have concentrated our study around the peak of the $\Delta (1232)$ resonance
region, i.e., when the energy of the virtual photon is equal to 340 MeV.

We have performed calculations for  $^{12} C$ and
$^{40} Ca$ at electron energies accessible at
Mainz and TJNAF.

\section{Virtual Compton Scattering in nuclei.}

The differential cross section for the VCS in nuclei (see Fig. 2) in the 
LAB system is given by:

\begin{equation}
\frac{d^5 \sigma}{d \Omega_{e'} d \Omega_{\gamma} d \omega'} =
\frac{m_e^2}{2 (2 \pi)^5} \,
\frac{M}{E'_A} \,
\frac{|\vec{k}'| \omega'}{|\vec{k}|} \,
\bar{\Sigma} \Sigma |{\cal M}|^2 
\end{equation}

\noindent
where, $k = (E, \vec{k})$ and $k' = (E', \vec{k})$ are the 
four-momenta of the incoming at outgoing electron respectively; 
$q = (\omega, \vec{q}\,)$ and $q' = (\omega', \vec{q}\,')$ are the four-momenta
of the intermediate and final photon respectively; $p = (M, \vec{0}\,)$ and
$p' = (E'_A, \vec{p}\,')$ are the momenta of the initial and final nucleus;
$m_e$ is the electron mass; $\cal{M}$ is the amplitude of the process. 
As we consider the unpolarized cross section,  we sum over final states and 
average over initial ones.

The amplitude ${\cal M}$ is given by:

\begin{equation}
{\cal M} = \frac{e}{q^2} \bar{u}_{r'} (k') \gamma^{\mu} u_r (k) 
\tilde{\Pi}_{\mu \nu}
(q , q') \varepsilon^{\nu} (q') 
\end{equation}

\noindent
in this expression $u_r (k) (u_{r'}(k'))$ are the Dirac spinors for the initial
(and final) electron with spin $r (r')$; $\varepsilon^{\nu} (q')$ is the 
vector polarization of the final photon; $\tilde{\Pi}_{\mu \nu} (q, q')$
is the hadronic current. In the delta region 
where we are interested in, and for spin saturated nuclei, this
hadronic part of the amplitude is dominated by the contribution of the
$\Delta (1232)$ \cite{PAR,CAR,RAF}, and it is given by \cite{CAR}:

\begin{equation} 
- i \tilde{\Pi}^{\mu \nu}_{\Delta} (q , q') =
\sum_{M_s} \sum_{M_I} (- i T^{\mu} (q)) (- i T^{\dagger \nu} (q')) \,
\int d^3 r \, e^{i (\vec{q} - \vec{q}\,') \cdot \vec{r}} 
\end{equation}

$$
\frac{1}{4}\frac{\rho (\vec{r}\,)}{\sqrt{s_{\Delta}} - m_{\Delta} + i 
\Frac{\tilde{\Gamma} (\sqrt{s_{\Delta}}, \rho (\vec{r}\,))}{2} -
\sum_{\Delta} (\sqrt{s_{\Delta}}, \rho (\vec{r}\,))}
$$

\noindent
in this expression, $\rho (\vec{r}\,)$ is the nuclear density,
 for which we use a two Fermi parameters parametrization \cite{NUC};
$\sqrt{s_{\Delta}}$ is the invariant mass of the nucleon-photon system.
In order to calculate this invariant mass, we assume for the initial nucleon
an average momentum given by:

\begin{equation}
\vec{p} = (\vec{q} \,' - \vec{q}\,) / 2
\end{equation}

\noindent
used also in \cite{PAR,PAV,CAR}. 
$\tilde{\Gamma} (\sqrt{s_{\Delta}}, \rho (\vec{r}\,))$ is the $\Delta (1232)$
width corrected by Pauli blocking \cite{WIDTH}.
$\Sigma_{\Delta} (\sqrt{s_{\Delta}}, \rho (\vec{r}\,))$ is the $\Delta$
self-energy which includes quasielastic, plus two-body and three-body contributions
 related to photon absorption\cite{WIDTH,SALCEDO}. 
$T^{\mu} (q)$ is the $\gamma N \Delta$ vertex, which, in the C. M. system
of reference is given by:

\begin{equation}
T^{\mu} (q) = \sqrt{\frac{2}{3}} \, 
\frac{f_{\gamma}}{m_\pi} \,
\frac{\sqrt{s_{\Delta}}}{m_{\Delta}} 
\left\{
\begin{array}{c}
  0  \\
\vec{S}  \times  \vec{q}
\end{array}
\right\}
\end{equation}

\noindent
where the factor $\sqrt{\frac{2}{3}}$ came from isospin, $f_{\gamma} = 0.12$
is the $\gamma N \Delta$ coupling constant \cite{RAF,BES}, and $\vec{S}$ is
the 1/2 to 3/2 transition spin operator normalized as:

\begin{equation}
< \frac{3}{2}, \, M \, |S^{\dagger}_{\nu}| \, \frac{1}{2}, \, m > =
{\cal C} \, \left(\frac{1}{2}\, 1\, \frac{3}{2}; m, \nu, M \right)
\end{equation}

\section{Background: Coherent Bremsstrahlung in nuclei.}

As we already said, photon emission from CB can be very important,
depending on the kinematics.
In this process the nucleus acts as a momentum source (with no energy
transfer) and the photon
is emitted from the electron. The cross section for this process in the
one photon exchange approximation is given by \cite{AMPARO}:

\begin{equation}
\frac{d^5 \omega}{d \Omega_{e'} \Omega_{\gamma} d \omega'} =
\frac{m^2_e}{2 (2 \pi)^5} \,
e^4 \, \frac{\omega' |\vec{k}'|}{|\vec{k}|} \,
L^{\mu \nu} A_{\mu} (\vec{q}\,) A_{\nu} (\vec{q}\,)
\end{equation}

\noindent
with $L^{\mu \nu}$
the leptonic tensor for the process
(detailed expressions can be found in ref. \cite{AMPARO});
$A_{\mu} (\vec{q}\,)$ the
Coulomb potential in momentum space, which is the Fourier transform of the
Coulomb potential in coordinate space,

\begin{equation}
A^{\mu} (\vec{x}) = 
(A^0 (\vec{x}), \vec{0}) = 
\left(
\frac{Z e}{4 \pi} \int d^3 x' \frac{\rho_{p} (x')}{|\vec{x} - \vec{x}'|} \, ,
\, \vec{0} \right).
\end{equation}

In order to calculate the total cross section for $A(e,e'\gamma)A$ 
we include also the interference between this process and the VCS.

\section{Real Compton Scattering.}

In this section we compare  VCS and RCS in
order to see if it is possible to extract the later from the first one.

The cross section for the RCS process (see Fig. 4) in the LAB system is given
by \cite{CAR}:

\begin{equation}
\left( \frac{d \sigma}{d \Omega_{\gamma}} \right)_{\hbox{real}} =
\frac{1}{16 \pi^2} \,
|{\cal M} (\vec{q}, \vec{q}\,') |^2.
\end{equation}

Once again, we are interested on the $\Delta$ resonance region. Then, the
amplitude ${\cal M}$ is given by:

\begin{equation}
{\cal M} (\vec{q}, \vec{q}\, ') = \tilde{\Pi}^{\mu \nu}_{\Delta} 
(\vec{q}, \vec{q}\,') \varepsilon_{\mu} (\vec{q}\,) \varepsilon_{\nu}
(\vec{q}\, ')
\end{equation}

Now, in order to compare this cross section to the VCS one,
we write down the cross section for unpolarized photon production
in (e,e') reactions \cite{ROH}:

$$
\frac{d^5 \sigma}{d \Omega_{e'} d \omega' d \Omega_{\gamma}} =
\Gamma \left\{
\frac{d \sigma_T}{d \Omega_{\gamma}} + \epsilon 
\frac{d \sigma_L}{d \Omega_{\gamma}} + \right.
$$

\begin{equation}
+ \epsilon \, \frac{d \sigma_{p}}{d \Omega_{\gamma}} \, 
cos 2 \phi_{\gamma} + \sqrt{2 \epsilon (1 + \epsilon)} \,
\frac{d \sigma_I}{d \Omega_{\gamma}} \, cos \phi_{\gamma} \left. \right\}
\end{equation}

\noindent 
where

\begin{equation}                     \label{gamma}
\Gamma = \frac{\alpha}{2 \pi^2} \frac{1}{- q^2} \frac{|\vec{k'}|}{|\vec{k}|}
\frac{1}{1 - \epsilon} k_{\gamma}
\end{equation}

$$
\epsilon = \left( 1 - \frac{2 \vec{q} \, ^{2}}{q^2} \tan^2 (\theta / 2)
\right)^{- 1}
$$

$$
k_{\gamma} = \frac{s - M^2}{2 M}
$$

\noindent
In the $\Delta (1232)$ region, the cross section for the VCS is dominated
by the transverse part \cite{PAV}:

\begin{equation}
\frac{d^5 \sigma}{d \Omega_{e'} d \omega' d \Omega_{\gamma}} 
\simeq
\Gamma \frac{d \sigma_T}{d \Omega_{\gamma}}
\end{equation}

\noindent
and for small angles between the final and initial electrons, the intermediate
virtual photon is almost real and then,

\begin{equation}
\frac{d \sigma_T}{d \Omega_{\gamma}} \simeq
\left( \frac{d \sigma}{d \Omega_{\gamma}} \right)_{\hbox{real}}
\end{equation}

Then, in principle, it is possible to extract the RCS cross section from the
VCS one. We will see in the results section how good this approximation is.
In order to compare both processes the energy of the incoming photon in the
RCS is taken as
 $q^0 = 340 \, MeV$ (the same value than in the VCS for the intermediate
photon), 
but then $q = q^0$ unlike for the VCS case.

\section{Results.}

\subsection{Virtual Compton scattering compared to the background (Coherent 
Bremsstrahlung).}

In Figs. 5-10 we show the differential cross section for VCS (dotted lines),
compared to the background (dashed lines), as well as the coherent sum of
both processes (solid lines). We have done such calculations for $^{12} C$
and $^{40} Ca$ for different angles between the final and initial electron,
$\theta_{e'}^{LAB}, (\theta_{e'}^{LAB}
= 2, 5, 10, 15$ degrees), and for different energies
of the incoming electron, $E_e, (E_e = 500, 800$ and $2000 \, MeV$), keeping 
always the energy of the intermediate photon equal to $340 \, MeV$, in order
to be always in the delta resonance region. We have plotted the differential
cross section as a function of the angle between the final photon and the
intermediate photon in the LAB system $(\theta^{LAB}_{\gamma})$. We have
integrated over the angle $\phi_{\gamma}$ of the photon.

As a first impression, we can see in Figs. 5-10 that the cross section for both
processes decreases
 when one increases the angle of the outgoing electron,
$\theta_{e'}^{LAB}$, and also when we increase the energy of the incoming electron,
$E_e$. Then, we can first conclude that the experimental study of the present
process would be favoured by small electron angles, $\theta_{e'}^{LAB}$,
and small
electron energies, $E_e$. 

In addition, we can see that the behaviour of the cross section is very 
similar for $^{12} C$ and $^{40} Ca$ except by the different nuclear form
factor, and the fact that cross sections depend on $A^2$, and then, the cross
section for $^{40} Ca$ is larger than for $^{12}C$.

Looking more carefully to the cross section, we can see that for 
$E_e = 500 \, MeV$, and for small electron 
angles $( \theta_{e'}^{LAB} = 2 - 5$ deg.) the VCS
dominates the cross section for photon angles larger than $25 - 30$ degrees.
For $\theta_{e'}^{LAB} = 10 - 15$ degrees we can see that it is
necessary to go to higher photon angles $(\theta^{LAB}_{\gamma} > 40 - 60$
deg.) in order to see the VCS signal over the background. Very similar 
features appear for $E_e = 800 \, MeV$, although here it is 
necessary to look at 
higher photon angles in order to measure the VCS.

 A different behaviour appears for larger electron energy
$(E_e = 2000$ $MeV)$.
In this case, we can see that for very small electron angles 
$\theta_{e'}^{LAB} = 2$ deg.) it is possible to see the VCS signal over the background 
for $\theta_{\gamma}^{LAB} > 25$ deg. However if we increase the electron angles it 
is necessary to go to very high photon angle $(\theta^{LAB}_{\gamma} > 120$
deg. for $\theta_{e'}^{LAB} = 15$ deg.) in order to measure the VCS.

\subsection{Virtual Compton Scattering compared to Real Compton Scattering.}

In Figs. 11-16 we can see the differential cross section for the RCS (dotted
lines) compared to the VCS cross section divided by the factor $\Gamma$ of Eq.
(\ref{gamma}) (solid line), and we have also plotted the transverse part of the
cross section, $d \sigma_T / d \Omega_{\gamma}$ (dashed lines).


The first thing that we can see is that the transverse part of the cross
sections for the VCS (dashed line) is almost the same than the total ones
(continuous lines): 
both lines, the continuous one and the dashed line, almost overlapping for all
energies.

Comparing now both the VCS and RCS cross sections at $E_e = 500 \, MeV$ we
see that they are essentially equal at $\theta_e < 15 $ deg.
If we move to larger electron energies,
$E_e = 800 \, MeV$ this agreement between both calculations remains
good for small
electron angles ($\theta^{LAB}_{e'} = 2 - 5$ deg.), but is not so good for
larger electron angles ($\theta^{LAB}_{e'} = 10 - 15$ deg.).

Finally, at $E_e = 2000 \, MeV$, only at very small 
electron angles, ($\theta^{LAB}_{e'} = 2 $ deg.), is there 
 agreement between the
VCS and RCS cross section and we observe important
discrepancies between
them for larger electron angles.

Thus, we see that in order to extract the cross section for the RCS from the
VCS process it is necessary to measure at small electron energies, and also
small electron angles. Even then one can not get the VCS cross sections at
angles of the photon smaller than 30 degrees since at small photon 
angles the Bremsstrahlung background dominates the cross section. This might
look like an important limitation of the method, but in practice it is not.
The reason is that the forward part of the RCS cross section can be
very well reproduced \cite{CAR,AUS}
   via the optical theorem, from experimental
data on the total photonuclear cross section \cite{AHREN,CARLOS}    
and the real part
of the RCS forward amplitude obtained from dispersion relations 
\cite{WEISE} .
Furthermore the fall down of the cross section up to 50-60 degrees is
well reproduced in terms of the nuclear form factor. Thus the genuine
new information contained in RCS lies precisely in the not very
forward angles \cite{SCHUM}, 
 where in fact there are still discrepancies between theory
and experiment \cite{PAR,KOC,CAR}.

	Given the limited amount of data on RCS, and the
 present persistent discrepancies between theory and experiment, 
obtaining more data on RCS is an important task. The method derived
here can make this goal easier than it has been so far.

\section{Conclusions}

Our conclusions can be summarized in two main points: 

On the first hand, we
have studied the VCS cross section in the $\Delta (1232)$ region. We have
compared this cross section with the background (coming from the
CB) that one would encounter in the implementation of the 
experiment. We have found the accessible  experimental regions. We have seen
that for small angles of the outgoing photon the CB dominates the
reaction. However, at  photon angles
around and above 30 deg.,
 it is possible to measure VCS.
In addition, we have seen that the most favourable
case happens for small electron energies and small outgoing electron 
angles.
In this case, the VCS cross section is maximum and, although the CB is also
maximum, the last one decreases faster than the first one with the outgoing
photon angle.

On the second hand, we have compared the VCS cross section with the RCS one.
We have seen that for small electron energies and small electron angles it is
possible to extract the RCS cross section from the measurement of the VCS
one. This is due to the fact that in such conditions ($\theta^{LAB}_{e'}$
small and $E_e$ small), the intermediate virtual photon in the VCS process
is almost real, and what we really have in this case is almost real Compton
scattering. The present study has, however, quantized how real the photon
must be  in order to have the VCS and RCS cross sections equal at level
of 2-3$\%$.
However, we should not forget about the already commented 
background which would appear in the measurement of VCS and which
makes the region of small angles not experimentally accessible. On the
other hand we argued that the experimental information on
forward Compton scattering is already contained in the existing precise
measurements of the total photonuclear cross section. Hence the region of 
angles which can be explored with the present method is precisely the
one offering genuine new information about RCS.
\vspace{0.5cm}
	
	We would like to acknowledge partial support from CICYT contract
number AEN 96-1719. One of us, A. Gil, wishes to thank the
Conselleria d'Educaci\'o de la Generalitat Valenciana for financial support.

\newpage

{\bf \underline{Figure captions:}}

\vspace{0.5cm}

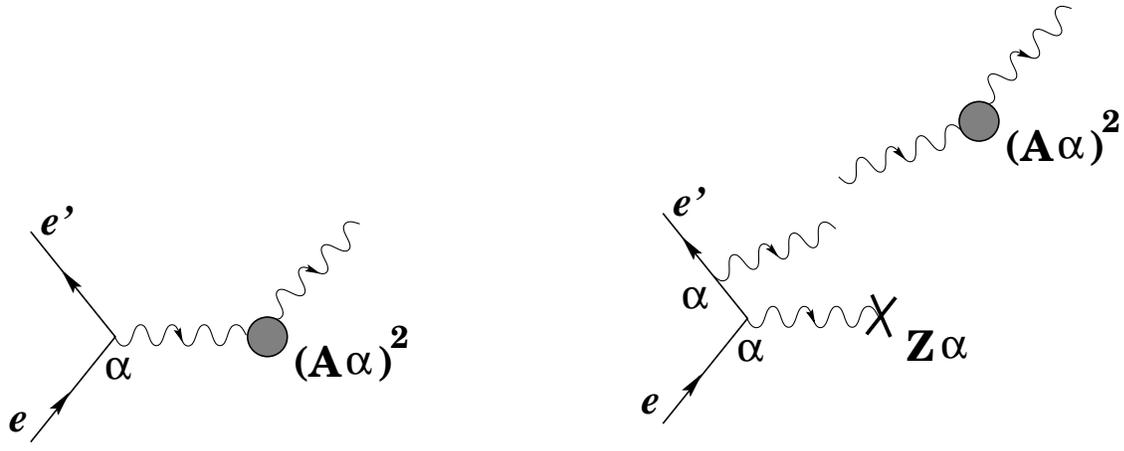
Figure 1: Schematical comparison between Real Compton Scattering and Virtual
Compton Scattering.

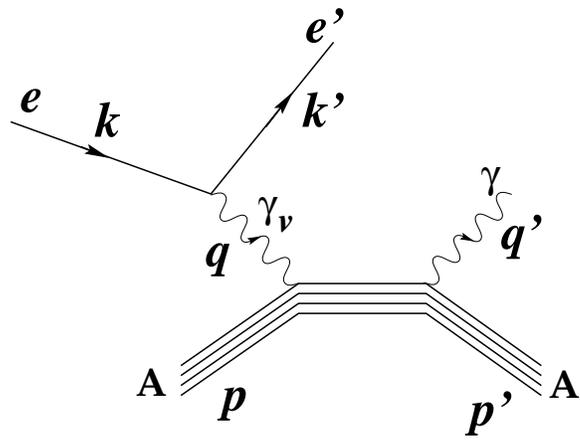
Figure 2: Diagrammatic representation of the VCS process in nuclei. In the
present case one of the hadronic lines in the intermediate states would be
a $\Delta(1232)$.

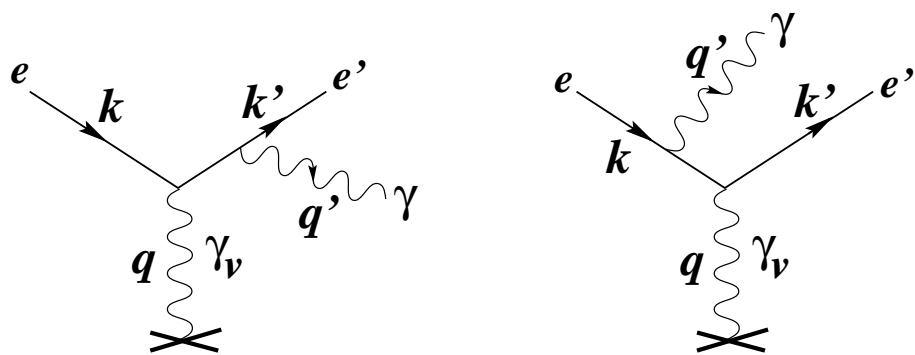
Figure 3: Feynman diagram for the coherent Bremsstrahlung in nuclei.

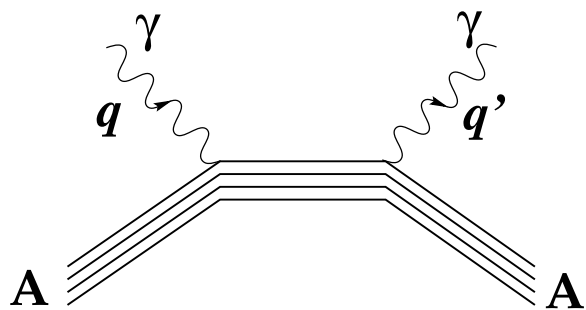
Figure 4: Diagrammatic representation of the RCS process in nuclei. In the
present case one of the hadronic lines in the intermediate states would be
a $\Delta(1232)$.

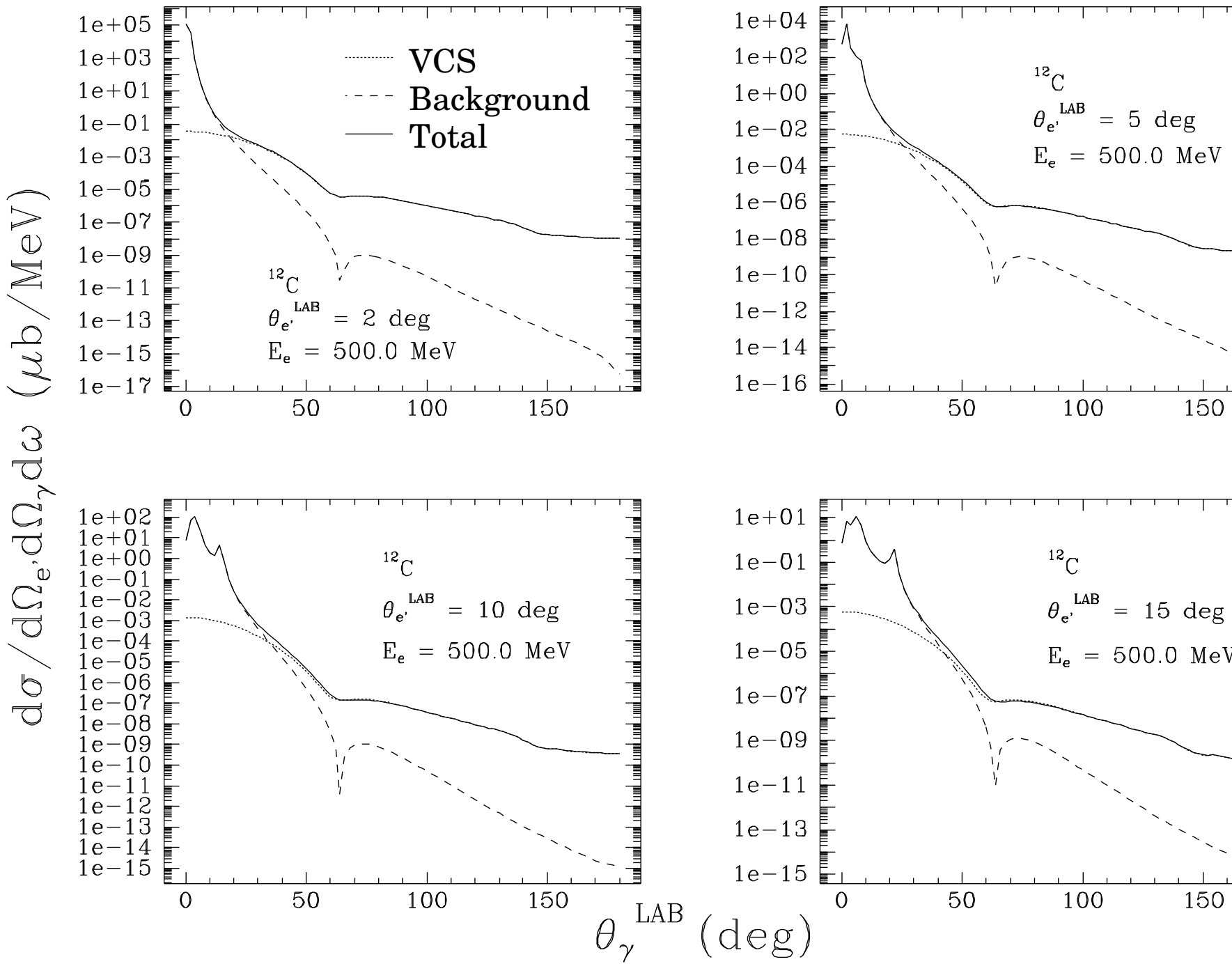
Figure 5: Differential cross section for the VCS process (dotted lines), 
Coherent Bremsstrahlung (dashed lines) and the total cross section (solid lines)
in $^{12} C$ for different outgoing electron angles 
($\theta^{LAB}_{e'} = 2, 5, 10$ and $15$ deg.) and for an incoming electron
energy, $E_e$, of $500 \, MeV$ as a function of the final photon angle,
$\theta^{LAB}_{\gamma}$
which is the angle between $\vec{q}$ and $\vec{q}\,'$. 
We always fix the transfer energy $E'_e - E_e$ equal
to $340 \, MeV$, in order to be around the delta resonance peak.

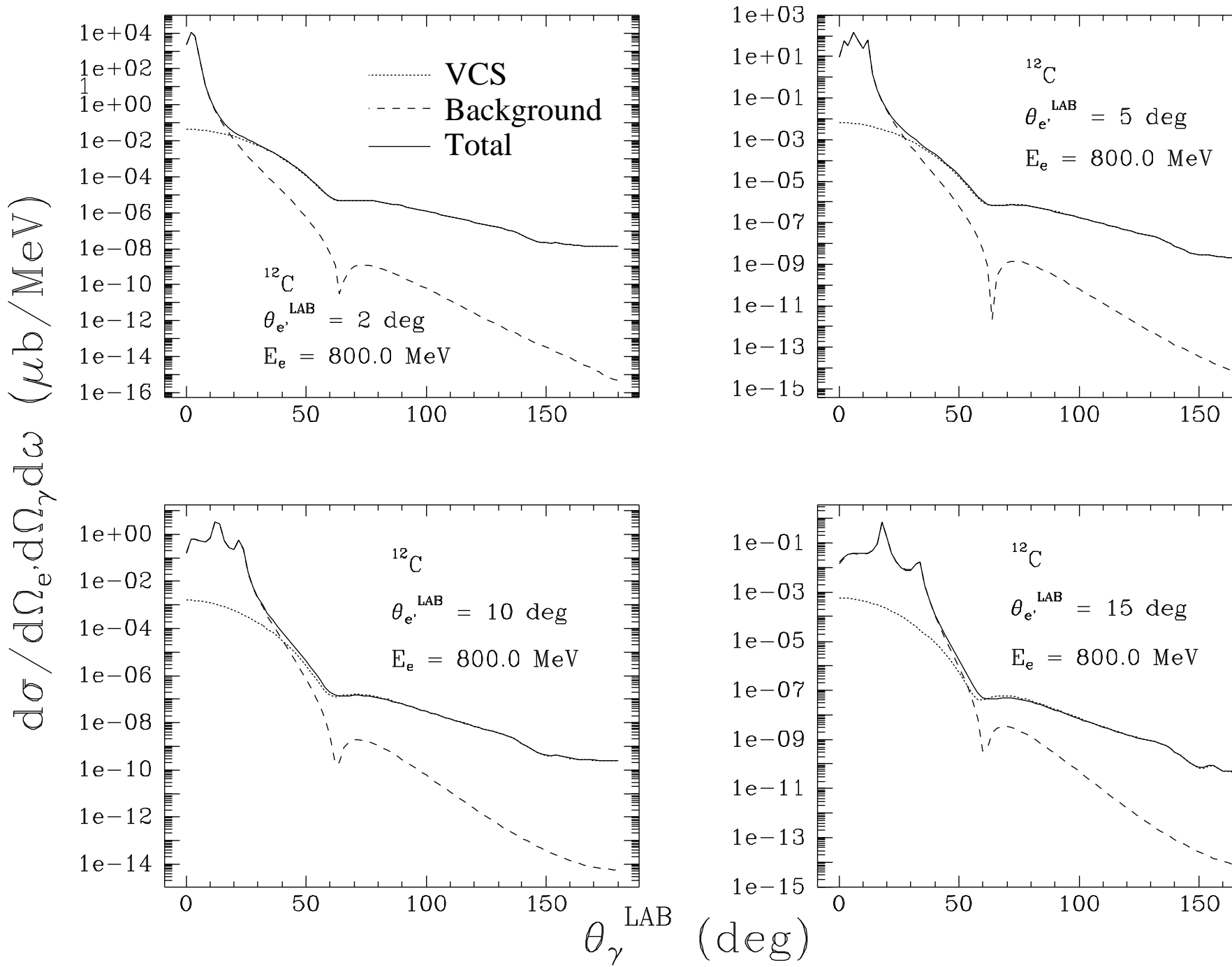
Figure 6: Same as Figure 5 for $E_e = 800 \, MeV$.

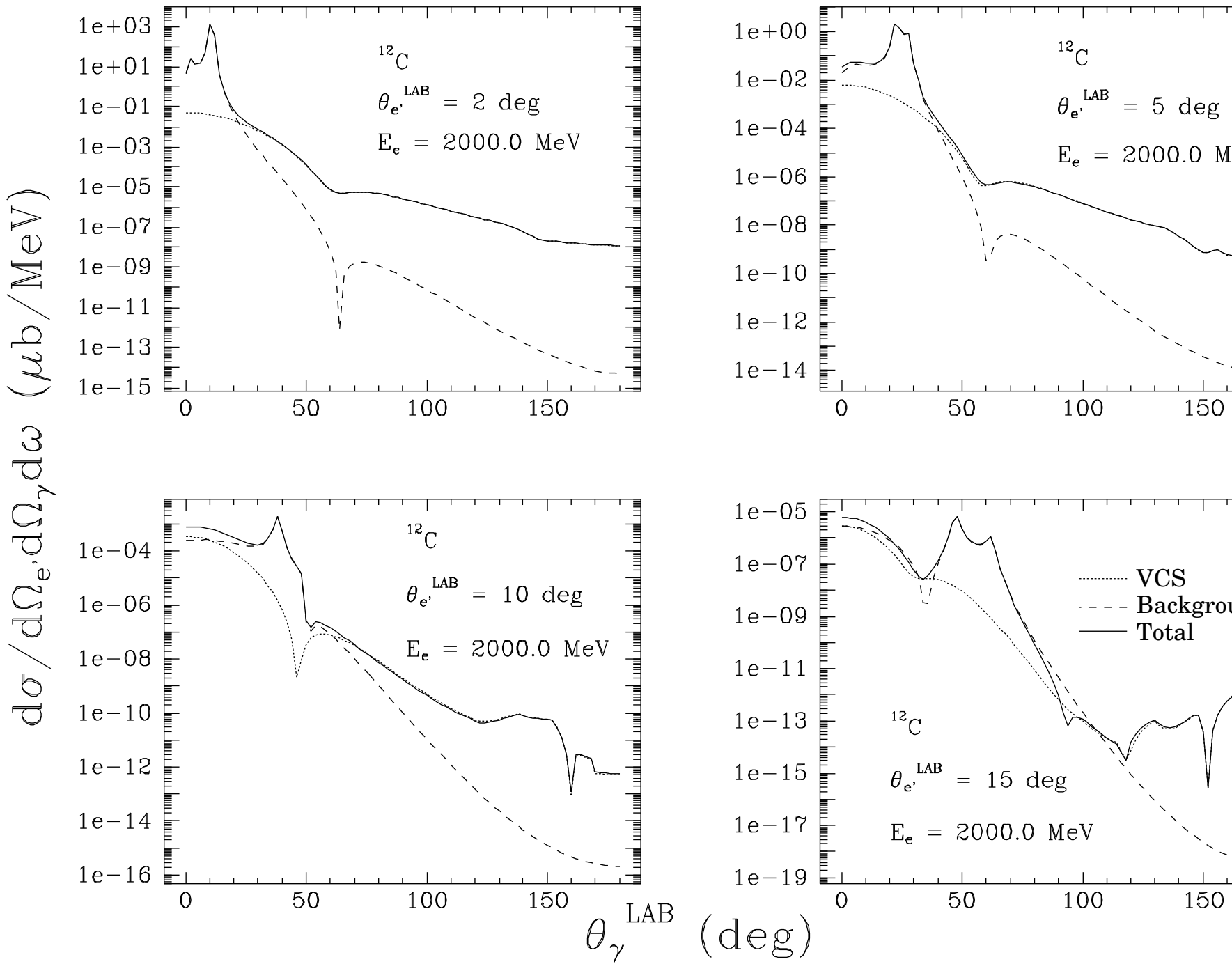
Figure 7: Same as Figure 5 for $E_e = 2000 \, MeV$.

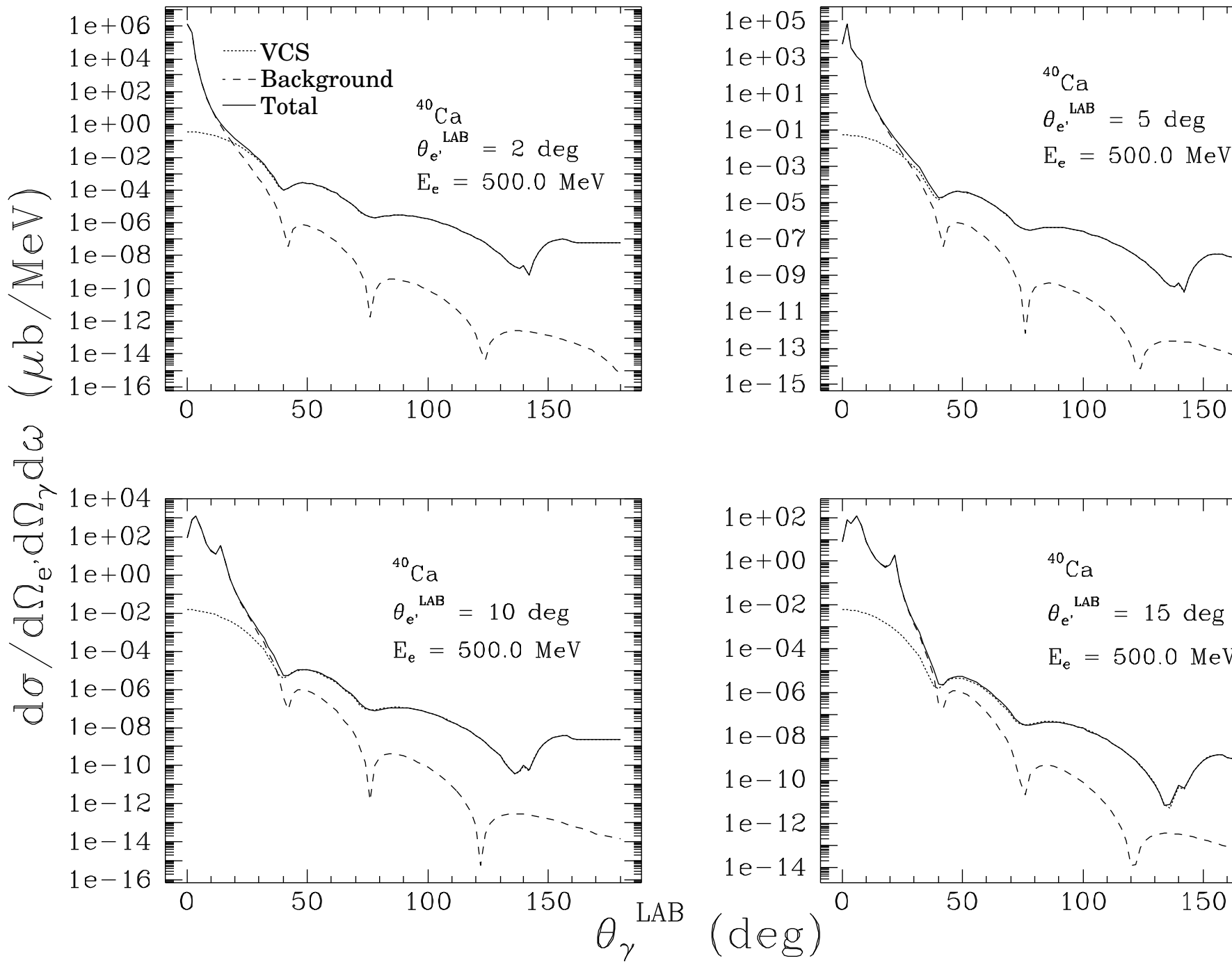
Figure 8: Same as Figure 5 for $^{40} Ca$.

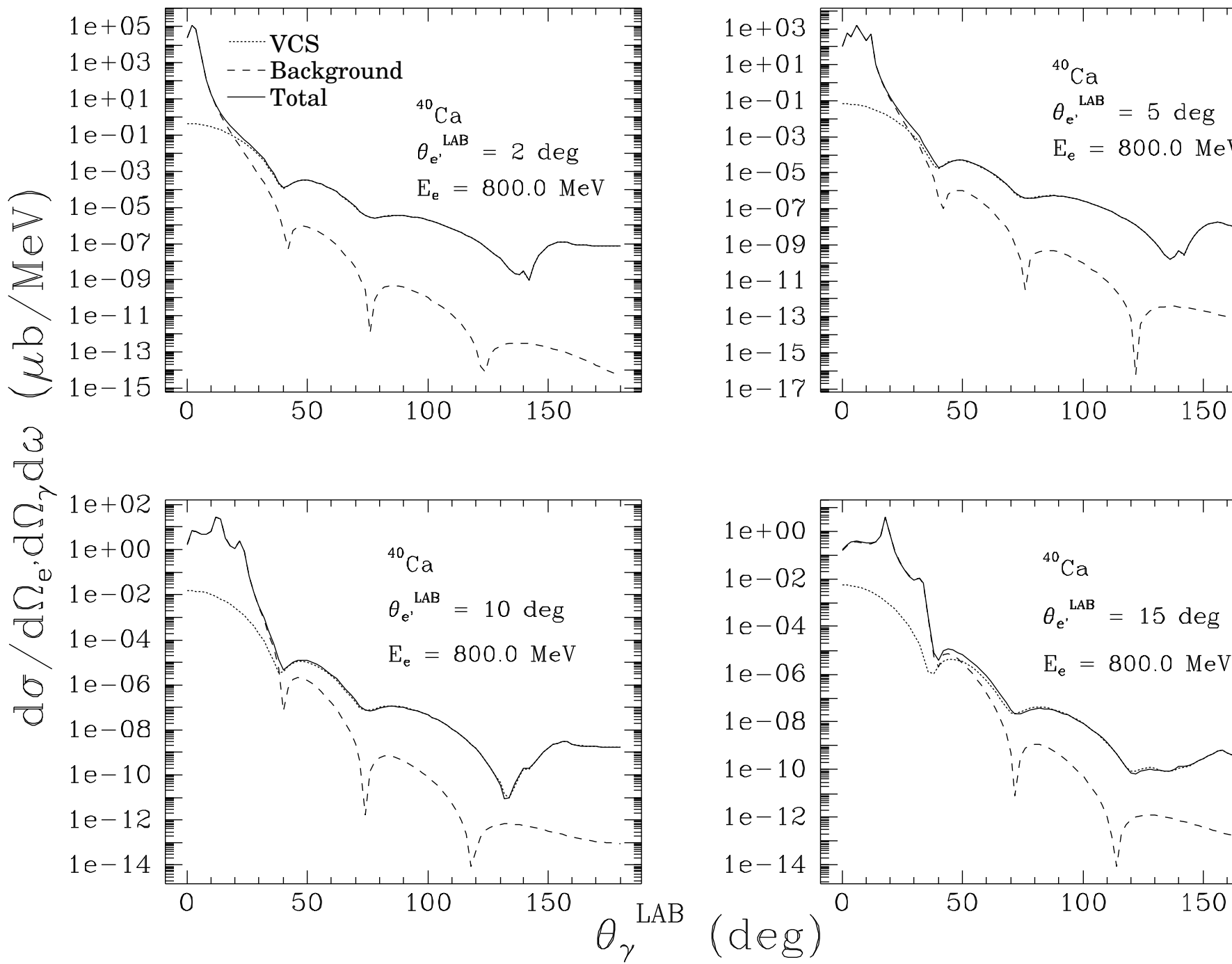
Figure 9: Same as Figure 5 for $^{40} Ca$ and $E_e = 800 \, MeV$.

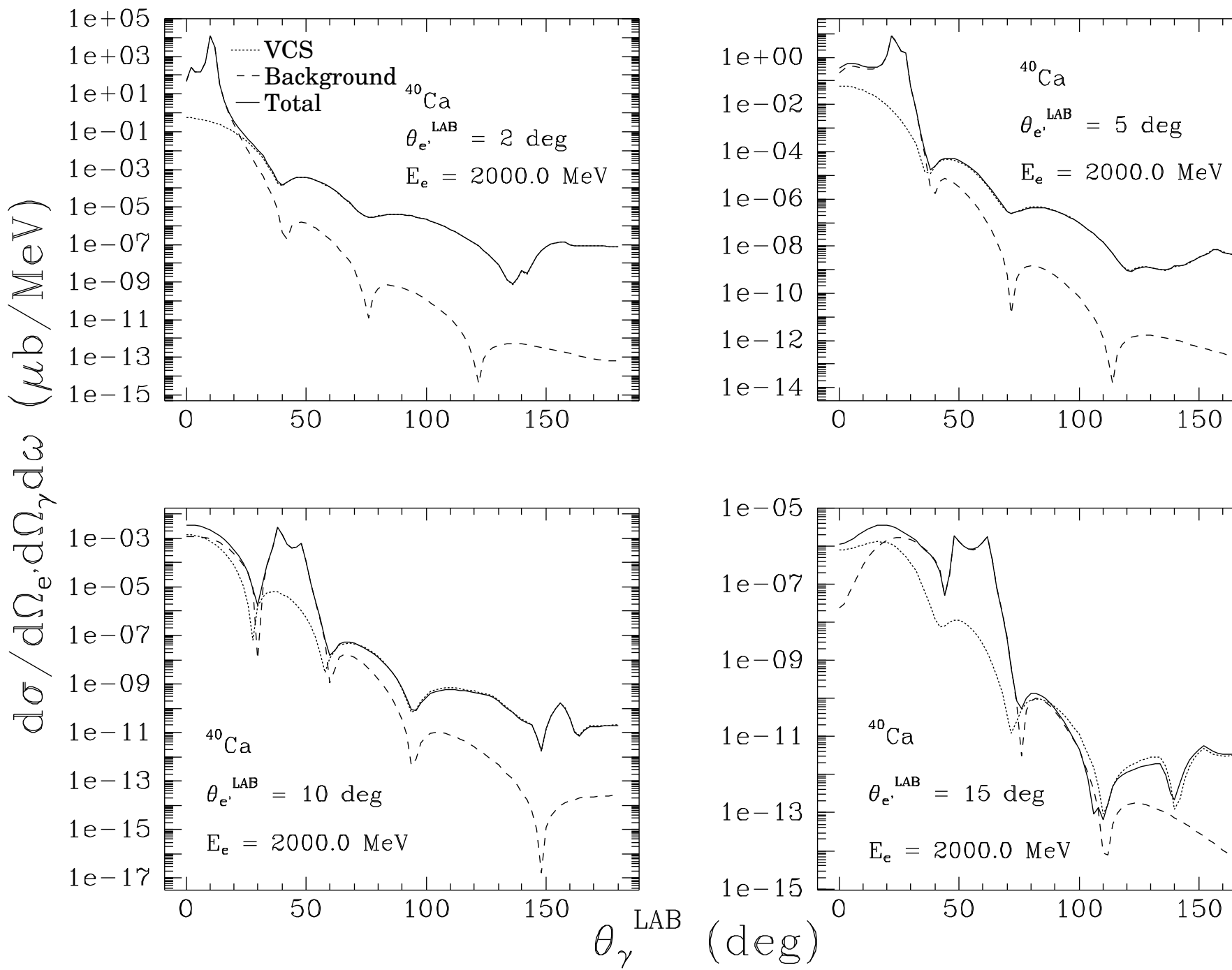
Figure 10: Same as Figure 5 for $^{40} Ca$ and $E_e = 2000 \, MeV$.

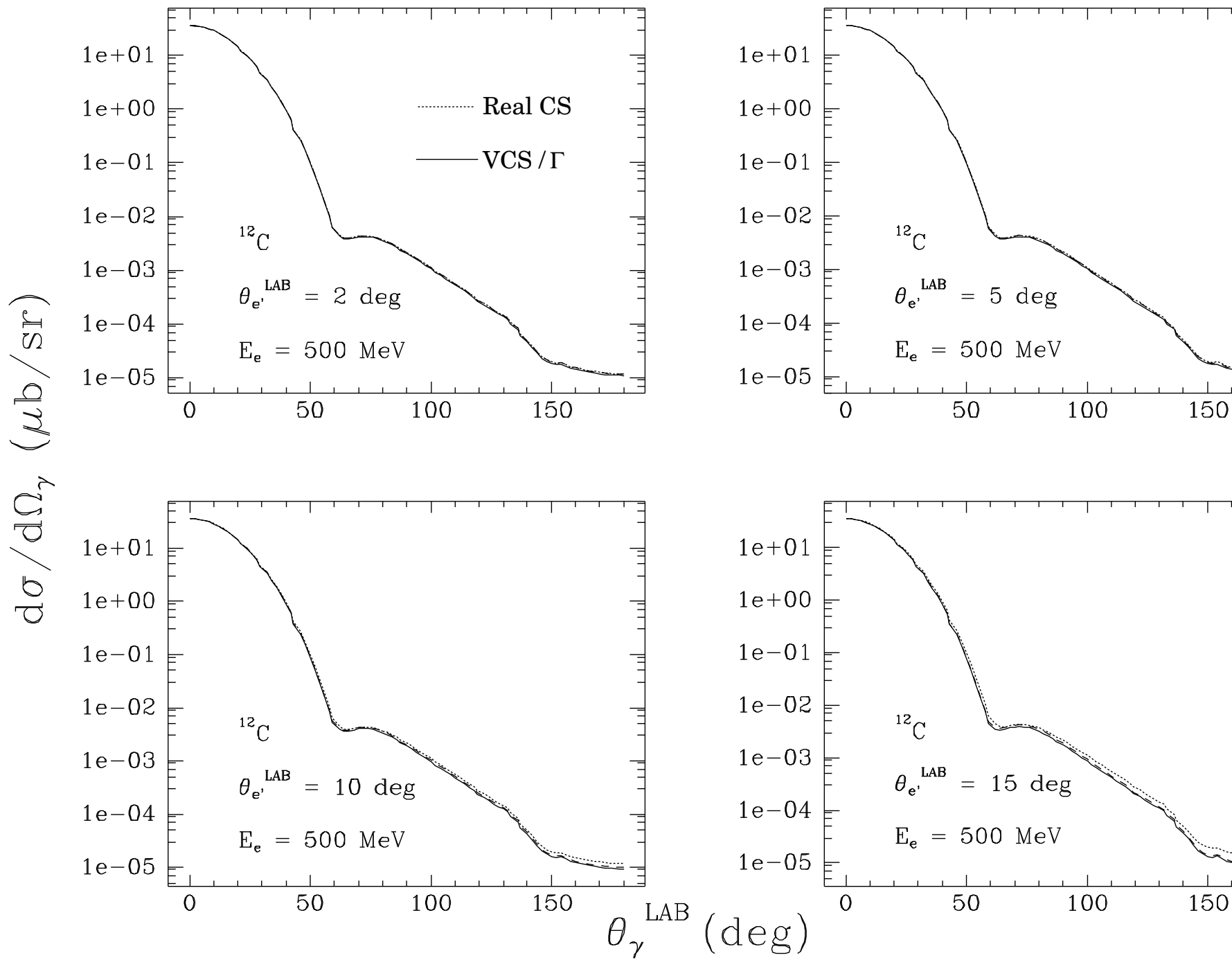
Figure 11: Differential cross section for the RCS process (dotted line), 
compared to the VCS divided by $\Gamma$ (solid line) and the transverse part
of the VCS cross section (dashed line) in $^{12} C$ for different outgoing
electron angles ($\theta^{LAB}_{e'} = 2, 5, 10$ and $15$ deg.) and for an
incoming electron energy of $500 \, MeV$ as a function of the final photon
angle, $\theta^{LAB}_{\gamma}$. We always fix the transfer energy
$E'_e - E_e$ equal to $340 \, MeV$ in order to be around
 the delta resonance peak.

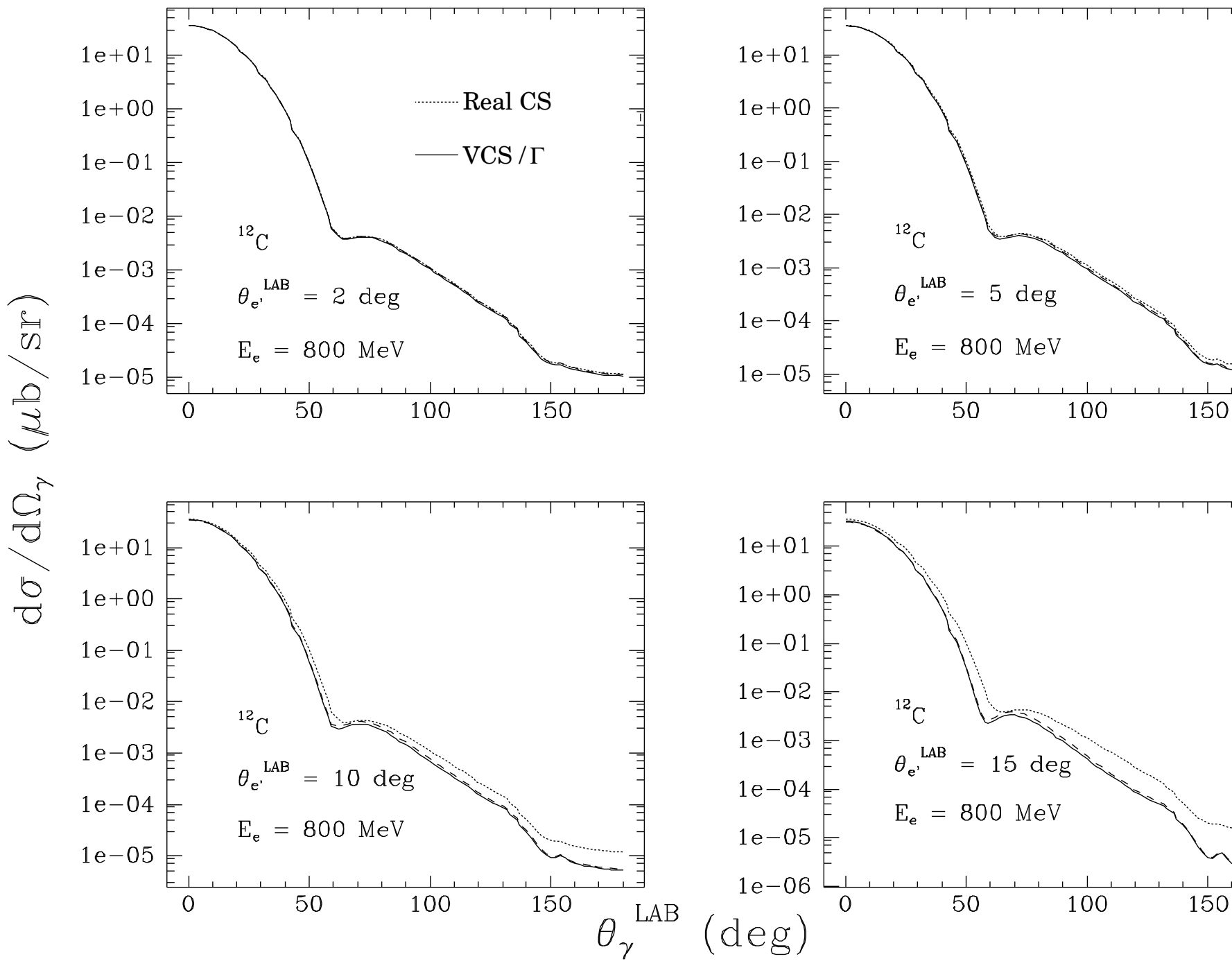
Figure 12: Same as Figure 11 for $E_e = 800 \, MeV$.

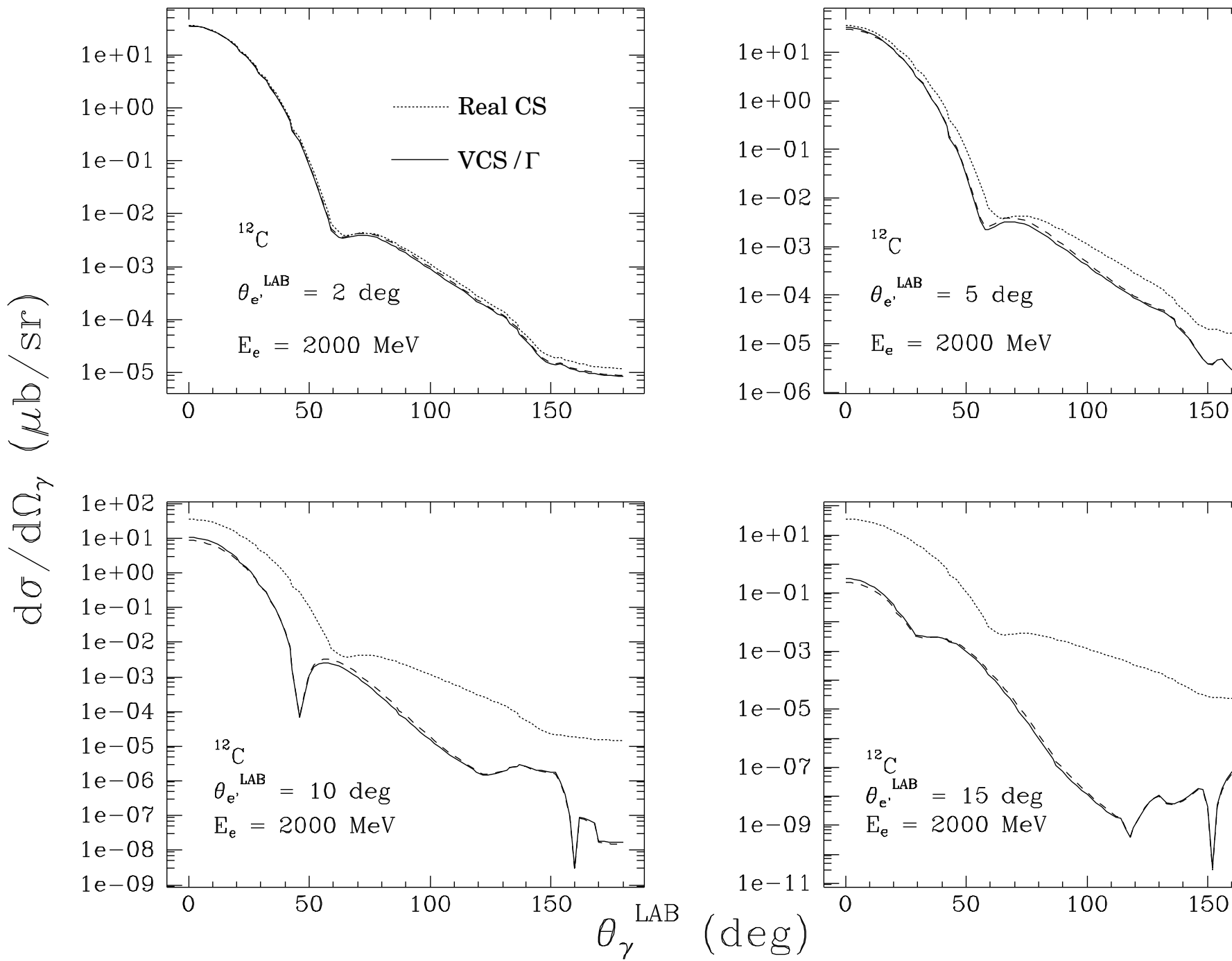
Figure 13: Same as Figure 11 for $E_e = 2000 \, MeV$.

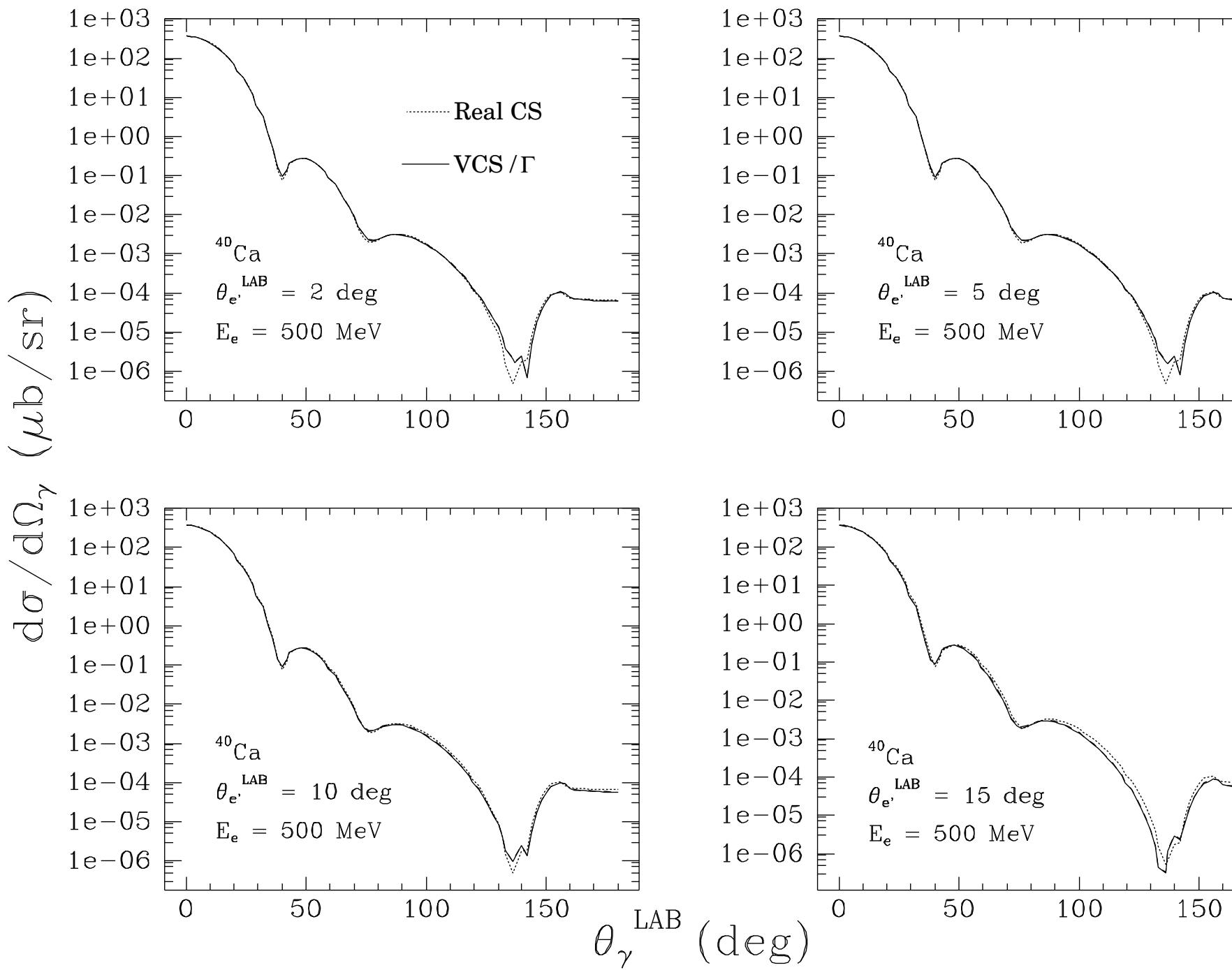
Figure 14: Same as Figure 11 for $^{40} Ca$.

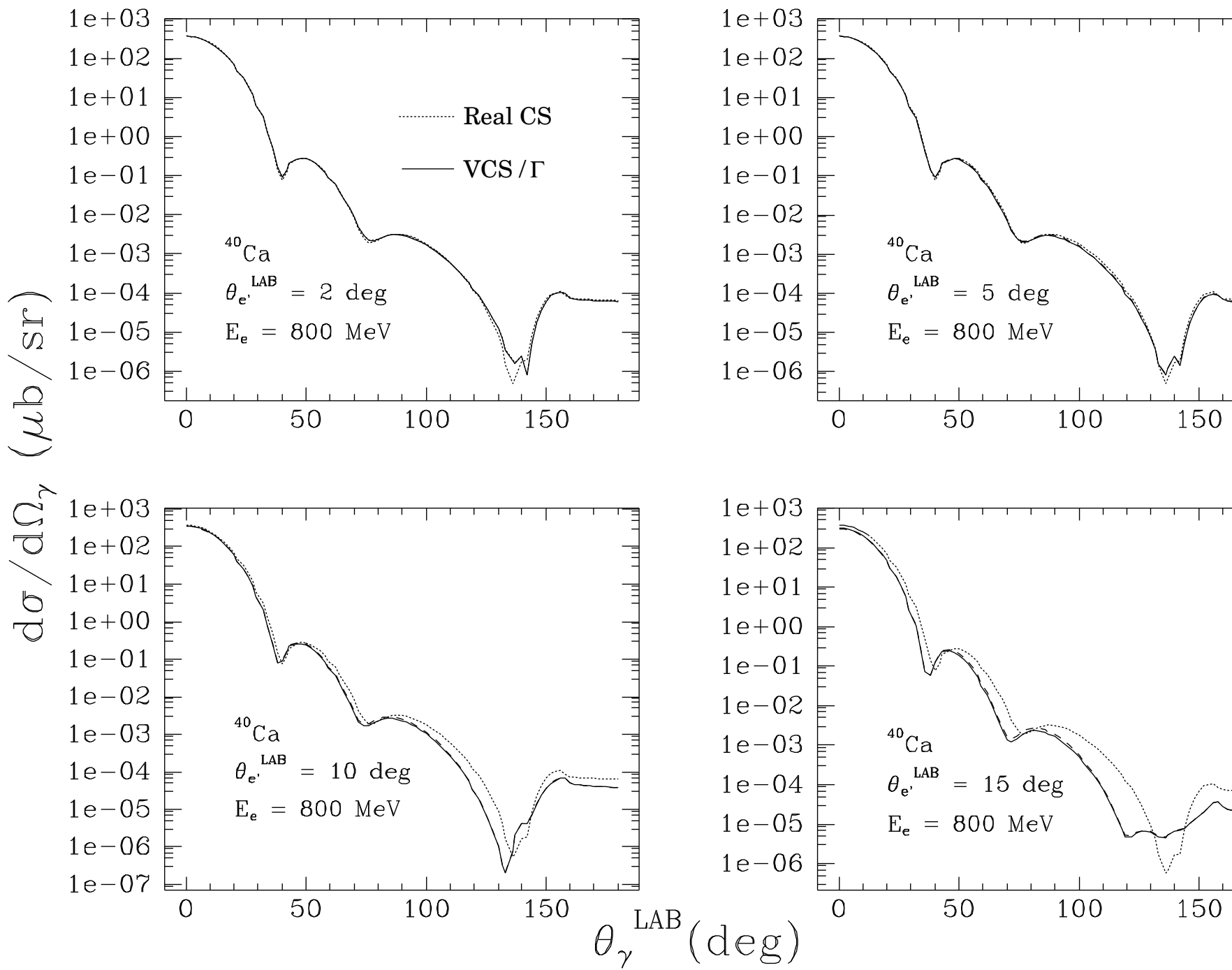
Figure 15: Same as Figure 11 for $^{40} Ca$ and $E_e = 800 \, MeV$.

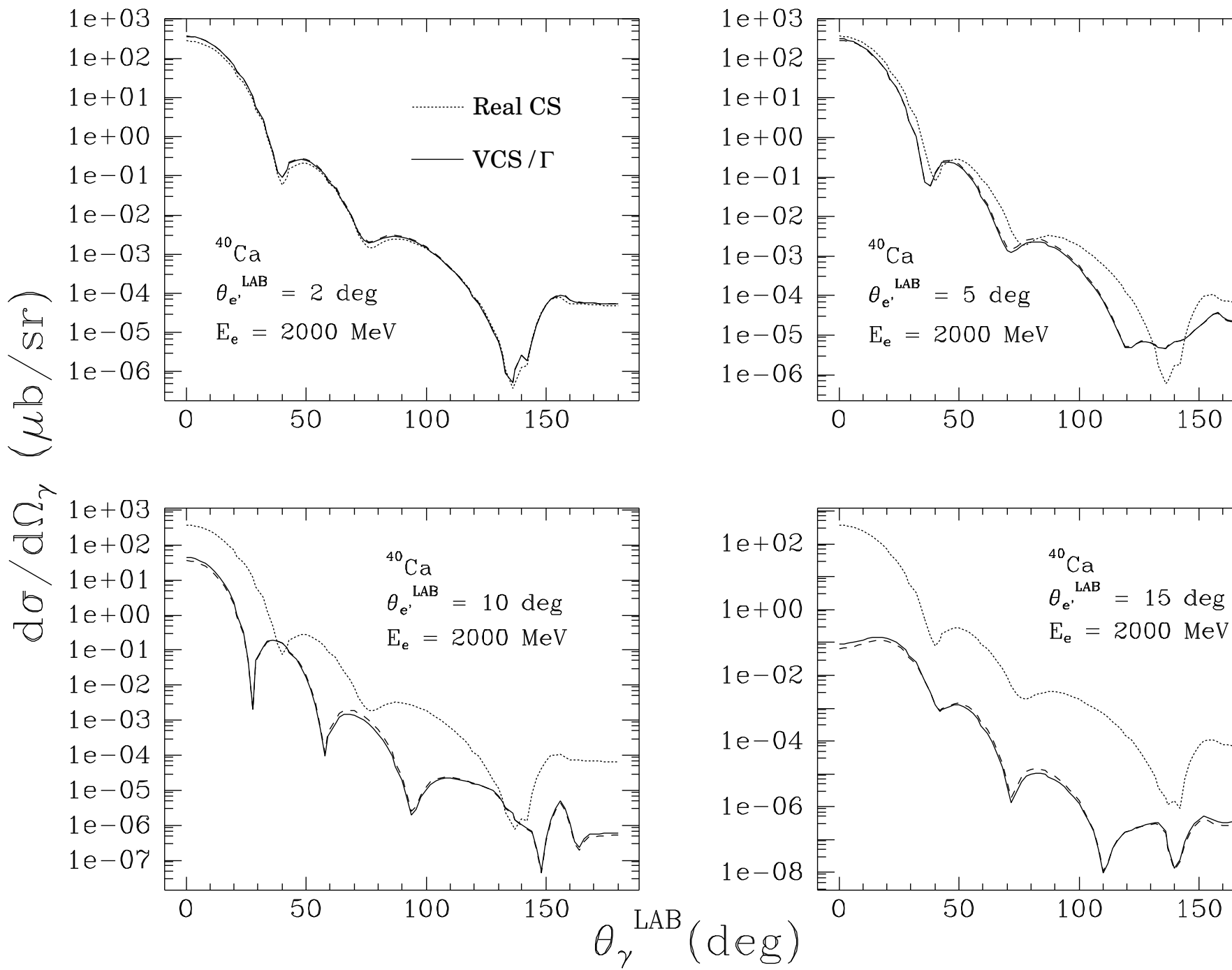
Figure 16: Same as Figure 12 for $^{40} Ca$ and $E_e = 2000 \, MeV$.

\end{document}